\documentclass[11pt]{article}

\usepackage[preprint]{acl}

\usepackage{times}
\usepackage{latexsym}

\usepackage[T1]{fontenc}

\usepackage[utf8]{inputenc}

\usepackage{microtype}

\usepackage{inconsolata}

\usepackage{graphicx}
\usepackage{booktabs}
\usepackage{pifont}
\usepackage{multirow}
\usepackage{hyperref}
\usepackage{xurl}
\usepackage{fontawesome5}

\makeatletter
\renewcommand{\@fnsymbol}[1]{%
  \ifcase#1\or\ensuremath{\dagger}\or\ensuremath{\ddagger}\or\ensuremath{\mathsection}\or\ensuremath{\mathparagraph}\or\ensuremath{\|}\or\ensuremath{**}\or\ensuremath{\dagger\dagger}\or\ensuremath{\ddagger\ddagger}\else\@ctrerr\fi}
\makeatother

\newcommand{\cmark}{\ding{51}}
\newcommand{\xmark}{\ding{55}}
\newcommand{\blu}[1]{\textcolor{black}{#1}}

\title{Evaluating Ambient Clinical Scribes in India:\\The Need for Multilingual Real-World Clinical Conversation Data}

\author{
  \textbf{Siddharth D Jaiswal\thanks{Work done while at Microsoft Research India.}\textsuperscript{1}},
  \textbf{Krithi Shailya\textsuperscript{$\dagger$2}},
  \textbf{Ashish Makani\textsuperscript{3}},
\\
  \textbf{Suvrankar Datta\textsuperscript{3}},
  \textbf{Sunayana Sitaram\textsuperscript{4}},
  \textbf{Mohit Jain\textsuperscript{4}}
\\
  \textsuperscript{1}VoiceArena,
  \textsuperscript{2} IIT Madras,
  \textsuperscript{3} Ashoka University,
  \textsuperscript{4}Microsoft Research India
}

\begin{document}
\maketitle
\begin{abstract}
Ambient clinical scribes (ACS) are being rapidly deployed at scale across Global South healthcare settings to reduce clinicians' documentation time, especially in overburdened environments such as India. These ACS are primarily developed or distilled from models built and validated on Global North speech, languages and consultation styles. Indian clinical encounters are brief, triadic, multilingual, code-mixed with low-resource languages, and conducted in highly resource-constrained, noisy settings -- increasing the likelihood of ASR and note-generation errors manyfold. We argue for a standardized evaluation infrastructure to assess whether these systems are safe, reliable, and well-suited to the Indian healthcare setting. We substantiate our claims through a mixed-methods study -- a systematic survey of publicly available patient-clinician conversational datasets, a quantitative comparison of some simulated datasets against conversational and cultural markers drawn from the Indian clinical-communication literature, and semi-structured interviews with four organizations building and deploying ACS in India. Our survey shows that there are no large-scale, publicly available, real-world benchmarks for ACS in India, with existing datasets being overwhelmingly synthetic or simulated. 
\blu{We note that amongst the available simulated datasets that we compare, the Global North datasets diverge significantly from the expected conversational and cultural structures of their Indian encounters.} Finally, our interviews reveal that deploying organizations have each built proprietary, incomparable evaluation pipelines, creating a fragmented ecosystem with no independent and reliable basis for procurement. We call for the development of a publicly shared, real-world, multilingual benchmark for ACS evaluation and outline the properties and policies such a benchmark would require. \footnote{Code: \url{https://github.com/v2dd-evals/ml4h-scribes-acs-findings-paper}} 
\end{abstract}

\section{Introduction}
\begin{figure}[!t]
    \centering
    \includegraphics[width=0.8\linewidth]{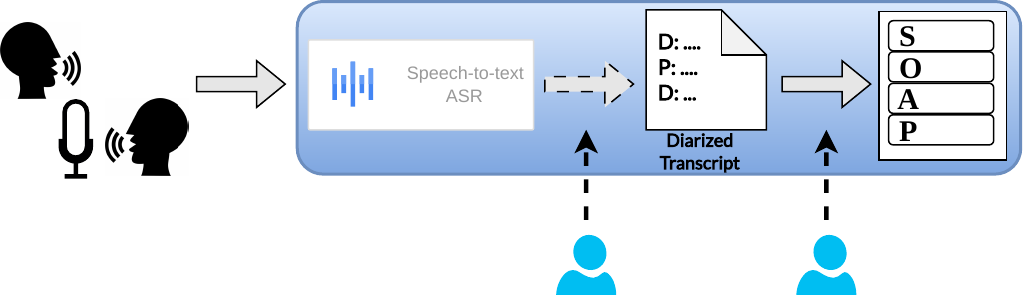}
    \caption{A standard ambient clinical scribe (ACS) pipeline. Patient–clinician audio is captured and transcribed by an ASR system into a diarized transcript, followed by a SOAP note. Humans may review and correct the transcript and note-generation. 
    }
    \label{fig:scribepipeline}
\end{figure}

Artificial Intelligence (AI) is increasingly used in diverse healthcare settings ~\citep{alowais2023revolutionizing,bohr2020rise}, applied across a range of clinical tasks such as imaging~\citep{panayides2020ai}, diagnosis~\citep{xu2024comprehensive}, and chat assistants~\citep{ramjee2025ashabot,ramjee2025cataractbot}. A recent but widely deployed application of AI is the ambient clinical scribe~\citep{tierney2024ambient, tierney2025ambient} (ACS) and related tools that aim to automate the documentation process~\citep{jain2025_ai_adoption_healthcare}. These systems record the patient-clinician encounter to automatically generate a structured clinical note, and have been observed to reduce documentation burden and clinician burnout~\citep{duggan2025clinician,olson2025use}. The output of these tools is passed to Electronic Health Record (EHR) systems to inform downstream analysis, prescriptions, billing, insurance, and subsequent clinical decisions (see Figure~\ref{fig:scribepipeline}).\\
The ambient clinical scribe market is dominated by organizations headquartered in the United States and Western Europe~\citep{microsoft_dragon_copilot,deepscribe_ai_medical_scribe,sunoh_ai,innovaccer_provider_copilot}. These tools are predominantly trained on clinical encounters drawn from Global North settings, and validated against benchmark data from the healthcare systems of those environments~\citep{korfiatis2022primock57,yim2023aci}. On the other hand, digital healthcare infrastructure is rapidly growing in the Global South, in countries such as India~\citep{mahajan2019artificial,das2024ai}, supported by government-led digital health initiatives such as the Ayushman Bharat Digital Mission~\citep{abdm_official}. With that, over the last two years in India, several ambient scribe systems have been piloted and deployed as a potential solution to the already-stressed medical workforce~\citep{mehta2024human} and to the documentation burden of managing hundreds of millions of patients~\citep{ekascribe_ai,augnito_omni,jatayu_voicedocai,plus91_digital_health,healthplix_halo}.

These deployment-contextual conditions differ substantially from those in the Global North in ways that directly affect how a clinical scribe performs. Indian consultations are routinely conducted across a long tail of low-resource languages~\citep{javed2022towards,bhogale2023vistaar}, and frequently code-mix within a single utterance~\citep{khanuja2020gluecos}, affecting both transcription and speaker attribution. Encounters are often triadic rather than dyadic, with family members and attendants speaking on the patient's behalf, making speaker attribution a non-trivial task. Interruptions from other medical staff (such as nurses and receptionists) are common. Consultations are shorter and denser, typically two minutes or less~\citep{irving2017international}. Local disease names and medicinal trade names are highly geography specific and may not be present in the vocabulary of Western models ~\citep{hasheminasab2026evaluating}. Building scribe systems for India’s complex medical environment is non-trivial, yet it remains unclear whether sufficiently representative training data and evaluation resources exist, with implications for reproducibility and informed procurement. 
Data infrastructure for clinical AI can take many forms, including resources for model development, validation, monitoring, and evaluation. Here, we focus specifically on evaluation infrastructure because it provides evidence for consequential decisions about system procurement, deployment, and governance.  
In a high-stakes domain such as healthcare, these decisions require a shared and credible basis for assessing capability, safety, and failure modes. While we focus specifically on evaluation resources, some of our findings are also relevant to training data curation.\\
We address the following research questions:\\
\noindent \textbf{RQ1:} What patient-clinician conversation datasets are currently available for the development and evaluation of ambient clinical scribes, and what are their key characteristics and limitations?\\
\noindent \textbf{RQ2:} To what extent do the characteristics of these datasets reflect the conversational and clinical realities of Indian healthcare settings?\\
\noindent \textbf{RQ3:} How are ambient scribes currently being developed, deployed, and evaluated by organizations operating in India?\\
Our study reveals the following key findings: First, there are no large-scale publicly available real-world benchmarks for ambient clinical scribes (ACS) deployed in India, in contrast to the growing body of such studies from the Global North~\citep{tierney2025ambient,tan2026impact,memon2026performance}. \blu{The only real-world Indian dataset~\citep{kumar2025asr} is not publicly available}. Our analysis shows that existing benchmarks are primarily synthetic or simulated and cover only a narrow set of languages, clinical specialties, regions, and tasks. Second, we identify a substantial gap between the cultural and interactional characteristics documented in Indian clinical communication research and those represented in available \blu{simulated} datasets. For instance, real Indian consultations open with terse, directive prompts (\textit{``Bataiye, kya hua?''} -- \textit{``Tell me, what happened?''}) that elicit trailing patient narratives with colloquial idioms (\textit{``Sir, sir phata ja raha hai\ldots''} -- \textit{``Sir, my head is splitting\ldots''}). Western consultations, by contrast, open with discursive invitations (\textit{``So $<$name$>$, I see you hurt your knee, tell me what happened''}), embed social exchange mid-consultation, and have fewer interruptions. Finally, qualitative analysis of semi-structured interviews with employees of organizations developing ACS for India reveals that each organization relies primarily on proprietary data and internal metrics for evaluation, rather than a shared, standardized benchmark.\\
The consequence of these findings is that hospitals and clinicians procuring these ACS systems have no baseline against which to compare vendors, regulators have no reference point for establishing standards, and systematic failures may remain indistinguishable from isolated errors. \\
\noindent \textbf{Contributions:} The key contributions of our work: (i)~a comprehensive survey of existing patient-clinician conversation datasets; (ii)~an analysis of the identified \blu{simulated} datasets characterizing their languages, settings, provenance, and construction, and identifying the gaps in these works; and (iii)~semi-structured interviews with four organizations building or deploying ambient scribes in India to surface the evaluation practices and constraints of current deployment.
\section{Background}
\subsection{The ambient clinical scribe pipeline}
\label{sec:pipeline}

An ambient clinical scribe (ACS) pipeline comprises two major components (Fig.~\ref{fig:scribepipeline}). The first is an Automatic Speech Recognition (ASR)  system that transcribes and diarizes the recorded patient--clinician audio so that patient and clinician-authored utterances can be distinguished in the resulting transcript~\citep{quiroz2019challenges}. Recently, large domain-adapted speech foundation models are being used to handle accented, overlapping, and code-mixed multi-lingual speech~\citep{kanaparthy2025realworld}. The second component is a note-generation system, most commonly an LLM, that converts the diarized transcript to a structured clinical note -- typically in SOAP (Subjective, Objective, Assessment, Plan) format. Often, this note is also fed to an EMR or HMIS system for further bookkeeping and analysis~\citep{quiroz2019challenges,kanaparthy2025realworld}. As the note-generation stage primarily uses the ASR output, transcript quality effectively determines note quality. Any errors in the first component compound in the second~\citep{tierney2024ambient}.

Each stage of the pipeline has unique errors. For example, at the ASR stage, diarization errors can misattribute a patient's reported symptom to the clinician, thus converting the complaint to a finding, or vice versa~\citep{rabotin2026propagation}. On the other hand, word-level or character-level errors can affect medication names, units, or quantities, resulting in clinically incorrect entities~\citep{anderson2025evaluating}. This is most consequential for dosage quantities, since audio-only transcriptions lack visual confirmation of the drug name, packaging, or quantity~\citep{menz2026vision}. Wrong-dose and monitoring errors are already a leading contributor to serious and life-threatening outcomes, even without automated systems as part of the loop~\citep{kale2012adverse}.

At the note-generation stage, errors further compound. \cite{anderson2025evaluating} report that transcription errors, when propagated to generated notes, are high enough to pose a moderate-to-severe risk to patients. \cite{bergman2026mind} found omissions in ASR transcriptions while \cite{kernberg2024using} performed a comparative study of LLM-generated SOAP notes and found that omissions accounted for the large majority of errors, followed by hallucinated additions and factual inaccuracies. Recent work by~\cite{koenecke2025perspective} argues that these errors disproportionately impact speakers with speech or language disorders, psychiatric presentations, and non-native accents. They call for extensive audits using diverse datasets and metric suites that go beyond aggregate word error rate. This concern is highly relevant to a linguistically diverse population such as India. With a substantially imbalanced patient-to-clinician ratio in the Indian healthcare system, mistranscriptions and SOAP note errors are easy for time-pressured clinicians to miss.

\subsection{ACS deployment in India}
ACS are extensively deployed throughout India. For example, reports~\citep{digitalhealthnews2025ekascribe,digitalhealthnews2024halo} have noted that tools like EkaScribe~\citep{ekascribe_ai} and H.A.L.O~\citep{healthplix_halo} have been rapidly adopted across clinics, hospitals, and healthcare chains, with early adopters reporting reduced documentation time. While remarkable, this uptake has occurred without any independent, standardized evaluation infrastructure, leaving health systems to rely on vendors' self-reported metrics and figures for their procurement decisions. These deployments are different from Global North contexts in the following structural ways.

Firstly, clinical encounters in India are high-volume and brief. India's doctor-to-population ratio remains substantially below WHO guidance, and in government hospitals, OPD (outpatient departments) clinicians routinely handle more than 200 patient visits per day, with an average consultation time of around 2 minutes~\citep{pandey2019patient,irving2017international,mehta2024human}, compared to approximately 18 minutes in the Global North~\citep{neprash2021measuring}. 

Second, clinical consultations in India are frequently triadic rather than dyadic. Family members routinely speak on the patient's behalf, negotiate with the clinician, and mediate disclosure of diagnosis and prognosis. Often, caregivers play a primary role in relaying symptoms and participating in treatment decisions~\citep{chawak2022treatment}. As most ASR systems are trained on dyadic Western consultations, the presence of an additional speaker becomes a challenging edge case and increases the risk of misattribution.

Finally, Indian healthcare infrastructure is markedly different from the well-sanitised settings of the Global North--consultations take place in noisy, crowded OPDs with regular interruptions from clinical staff, captured on shared devices with intermittent connectivity, over code-mixed, multilingual speech~\citep{javed2022towards,khanuja2020gluecos,bhogale2023vistaar}. These three structural features show that ACS systems deployed in India are operating in a substantially different setting, and therefore, standardized independent evaluations must be performed that reflect these conditions.

\begin{table*}[!t]
\centering
\scriptsize
\setlength{\tabcolsep}{1pt}
\caption{Characteristics of patient-clinician conversation datasets identified in our literature survey. Provenance distinguishes online medical interactions, simulated consultations, expert-generated data, and real clinical encounters. \faFileAudio[regular]: audio files, \faIcon*[regular]{file}: textual files, N.A.: information not available.}
\begin{tabular}{llrrcccc}%
\toprule
\textbf{Dataset} & \textbf{Source of data}  & \textbf{Lang (Size)} & \textbf{Modality} & \textbf{Provenance} & \textbf{Specialties} & \textbf{Public}\\ 
\midrule
Primock57~\citep{korfiatis2022primock57} & Mock Clinicians + Actors   & en (57) & \faFileAudio[regular] + \faIcon*[regular]{file} & Simulated & 12 & \cmark \\
ACI-Bench~\citep{yim2023aci} & Physicians + Actors & en (207) & \faIcon*[regular]{file} & Simluated & N.A. & \cmark \\

MediTOD~\citep{saley2024meditod} & \citet{fareez2022dataset} & en (213) & \faIcon*[regular]{file} & Simulated & 2 & \cmark \\

OSCE~\citep{fareez2022dataset} & Physicians + Med students   & en (273) & \faFileAudio[regular] + \faIcon*[regular]{file} & Simulated & 5 & \cmark \\

MTSDialog~\citep{abacha2023empirical} & MTSamples (Simulated)   & en (1.7K) & \faIcon*[regular]{file} & Simulated & 6 & \cmark \\

\citep{figure8dataset} & Kaggle & en (6.6K) & \faFileAudio[regular] + \faIcon*[regular]{file} & N.A. & N.A. & \cmark \\

MedDialog~\citep{zeng2020meddialog} & Web medical forum & en (250K), zh (3.4M) & \faIcon*[regular]{file} & Real online & 172 (en), 96 (zh) & \cmark \\


MedDG~\citep{liu2022meddg} & Web medical forum   & zh (17.8K) & \faIcon*[regular]{file} & Real online & 1 & \cmark \\

IMCS-21~\citep{chen2023benchmark} & Web medical forum   & zh (4.1K) & \faIcon*[regular]{file} & Real online & 1 & \cmark \\

ReMeDi~\citep{yan2022remedi} & Web medical forum   & zh (96.9K) & \faIcon*[regular]{file} & Real online & 843 & \cmark\\


NoteGen~\citep{eka_notegen} & Eka Healthcare & en, hi, mr (156) & \faIcon*[regular]{file} & Simulated & N.A. & \cmark \\

\citet{kumar2025asr} & Tertiary teaching hospital & en, hi, kn (162) & \faFileAudio[regular] + \faIcon*[regular]{file} & Real clinical & 2 & \xmark\\

ASR Notes~\citep{eka_asr} & Eka Healthcare & en (3.6K), hi (320) & \faFileAudio[regular] + \faIcon*[regular]{file} & Simulated & N.A. & \cmark \\
\bottomrule
\end{tabular}
\label{tab:litsurvey}
\end{table*}

\section{Experimental Design and Methods}
We use a mixed-methods study to characterize existing benchmarks and industrial products available for ACS in India. Our methodology combines (1)~a systematic literature survey of patient-clinician conversation datasets, (2)~a quantitative analysis of the identified datasets against a set of conversational and cultural markers relevant to Indian clinical encounters, and (3)~semi-structured interviews with organizations developing/deploying ACS in India. These three components address complementary aspects: publicly available benchmarks, whether these benchmarks reflect the conditions relevant to Indian healthcare settings, and how organizations deploying these systems design and evaluate their products.

\noindent \textbf{Dataset Survey:} We conducted a comprehensive literature review to identify candidate datasets containing patient-clinician conversations. 
We conducted a structured search across arXiv, Google Scholar, the ACL Anthology, and Hugging Face Datasets, using search terms including \textit{``doctor-patient dialogue dataset'', ``clinician-patient medical consultation'', ``medical conversation dataset''}, etc., and filtered the datasets containing patient-clinician interactions relevant to ambient scribing, clinical documentation or information extraction. 
For each dataset, we collected the following metadata: source, language, size, modality, real versus simulated or synthetic provenance, number of medical specialties, tasks, evaluation metrics, and whether the data is publicly accessible.\\
\noindent \textbf{Cultural and Conversational Marker Analysis}: To assess whether existing datasets can be used directly or translated to create a meaningful evaluation suite for Indian ACS systems, we developed a taxonomy of cultural and conversational markers from the literature on patient-clinician communication in India. We use a clinician-validated automated pipeline to assess whether a dataset contains the conversational markers typically found in Indian healthcare settings. The taxonomy covers several dimensions of clinical interaction, including conversation dynamics, language practices, interactional norms, and culturally situated descriptions. For each dataset, we compared its characteristics against the identified conversational and cultural markers and used an LLM-as-a-judge to assess the level of representation \blu{(further details are in Appendix~\ref{appendix:quant})}. This comparison allows us to identify differences in the underlying conversational structure; for example, a dataset may include conversations in Hindi, but may still not be culturally representative.\\
\textbf{Practitioner Interviews}: To understand how ACS systems are developed, deployed, and evaluated in practice, we conducted semi-structured interviews with senior employees from four organisations that build or deploy ACS systems in India. The organisations varied in product architecture, target users, deployment settings, scale, and degree of integration with electronic medical record (EMR) or hospital information management systems (HMIS), providing perspectives across multiple stages of commercial deployment. All interviews followed a common set of questions covering the system's primary use cases, input modalities and supported languages, handling of code-mixed speech and multiple overlapping speakers, data sources and collection practices, AI models and pipelines used, output formats and clinical coding, EMR/HMIS integration, deployment scale and settings, data privacy, storage and ownership protocols, user feedback mechanisms, evaluation metrics and datasets, and limitations and future roadmap. The interviews were semi-structured, lasting 45-60 minutes, allowing interviewees to elaborate on their responses. All interviews were conducted over Microsoft Teams and audio-recorded and transcribed. One of the researchers conducted all the interviews, while another researcher took notes and intervened only if needed. One researcher analyzed the transcripts and applied deductive thematic analysis~\citep{braunclarke2006using} identifying the codes from the interview topic areas; within each topic, the researcher iteratively coded the transcripts to identify repeating patterns and points of divergence across organizations. Codes were tracked in a shared spreadsheet and progressively grouped into higher-level themes (e.g. reliance on proprietary evaluation data, absence of shared benchmarks). As coding was performed by a single researcher, we treat the resulting themes as a single systematic interpretive account of the interview data rather than as inter-rater-validated coding, consistent with a reflexive thematic analysis stance~\citep{braunclarke2019reflecting}.

\section{Patient-Clinician Conversational Dataset Survey}
Table~\ref{tab:litsurvey} summarizes the datasets identified in our literature survey along key dimensions relevant to ACS evaluation, including source, provenance, language, dataset size, modality, and specialty coverage.
We observe that the resulting landscape is heterogeneous. Datasets range from small collections of transcribed consultations~\citep{yim2023aci,saley2024meditod,eka_notegen} to large-scale text corpora collected from online medical forums~\citep{yan2022remedi,zeng2020meddialog,liu2022meddg,chen2023benchmark}, with substantial diversity in their provenance, number of specialties, and intended use. MedDialog~\citep{zeng2020meddialog} is the largest resource we identify; however, the interactions are collected from web-based medical forums, where they are asynchronous and therefore not fully representative of live patient-clinician encounters. Other large Chinese-language resources are similarly derived from online medical interactions. These datasets provide substantial scale, but their modality and provenance limit their direct applicability to ACS.

\begin{table*}[!t]
\centering
\scriptsize
\setlength{\tabcolsep}{2pt}
\caption{Consultation characteristics averaged for each dataset. \textbf{C:P ratio}: clinician-to-patient word ratio; values below 1.0 indicate patient-dominant speech. \textbf{HT:Tx ratio}: history-taking turns per treatment turn; \textbf{Directiveness}: proportion of all clinician turns that are directive. \textbf{Clinical terms} is the proportion of turns containing medical or technical vocabulary. \textbf{Response-Agreement}: proportion of patient turns that agree without asking a follow-up question. \textbf{Interruptions} denote the proportion of turns containing interruptions.}
\label{tab:culturalmarkers}
\begin{tabular}{llcccccc}
\toprule
    \textbf{Dataset} & \textbf{Type} & \textbf{C:P ratio} & \textbf{HT:Tx ratio} & \textbf{Directiveness} & \textbf{Clinical terms} & \textbf{Response-Agreement} & \textbf{Interruptions} \\
    \midrule
    
    ACI-Bench (en) & Simulated & 2.06 & 1.5 & 20\% & 28\% & 18\% & $\sim$0\% \\
    
    MTS-Dialog (en) & Simulated & 1.95 & 1.4 & 18\% & 24\% & 21\% & $\sim$0\% \\
    
    Eka Care (hi) & Simulated & 0.89 & 4.2 & 62\% & 16\% & 64\% & 12\% \\
\bottomrule
\end{tabular}
\end{table*}

We identified only a small number of datasets containing real-world clinical conversations. \citet{kumar2025asr} claim to have collected a dataset of 162 audio files in English, Hindi, and Kannada for psychiatry-related consultations, intended primarily for automatic speech recognition rather than end-to-end scribe evaluation, \blu{but have not made it publicly available}.
\blu{Large real-world datasets such as MedDialog~\citep{zeng2020meddialog}, ReMeDi~\citep{yan2022remedi}, and MedDG~\citep{liu2022meddg} are curated from online medical forums and thus are not true transcripts. As noted in Figure~\ref{fig:scribepipeline}, an ACS must also perform an additional stage of note generation, which includes speaker attribution and clinical concept extraction. We did not identify any other large real-world clinical conversation dataset from India. To address the challenges of data privacy and sharing, clinicians' and/or patients' roles are often performed by trained actors who read from a script or respond to questions as if they were real patients; these simulated consultations are present in ~\cite{fareez2022dataset}'s dataset, MTSDialog~\citep{abacha2023empirical}, Primock57~\citep{korfiatis2022primock57} and ACI-Bench~\citep{yim2023aci}.}
Although these datasets are more closely aligned with an ACS, their small scale, limited language and accent coverage (only English spoken in the Global North), and reliance on simulated or expert-generated interactions constrain their relevance. 
The datasets open-sourced by Eka Care provide Indian-language resources for ASR~\citep{eka_asr} and note generation~\citep{eka_notegen}, but these are relatively small and task-specific, \blu{and are the only publicly available Indian-setting resources}. 
Our survey shows that no dataset, or combination of datasets, offers the scale, real-world provenance, multilingual coverage, conversational diversity, and task coverage required for ACS development and evaluation in India. The gap is particularly evident in the \textit{intersection of real clinical speech, Indian language diversity, and large-scale note-generation evaluation}. Most large datasets are text-based, in Chinese, and originate from online interactions; datasets closer to the ACS setting are substantially smaller and usually simulated; and the relatively few real-world Indian datasets are primarily designed for evaluating specific components of the pipeline, such as ASR.

\section{Do Existing Datasets Represent Indian Clinical Encounters?}

Next, we study whether the available \blu{simulated} datasets are representative of the conversational conditions encountered in Indian clinical practice. We first synthesized the patient-clinician communication literature into a set of cultural and conversational markers. These studies show Indian consultations are typically brief, clinician-directive, and power-asymmetric~\citep{doctorbehaviour, mehradoctorinfluence}, with patients offering colloquial symptom descriptions and heavy English--regional-language code-mixing~\citep{codemixing, idiomsofdistress}. The resulting markers span four angles: conversational control and participation (e.g., consultation duration and turn structure, interruptions~\citep{waitingtimedas}, family or attendant participation~\citep{attendantrole}); trust- and authority-driven patient deference and question-asking (e.g., confirmation and explanation practices~\citep{mehradoctorinfluence}, information-seeking or treatment-discussion patterns~\citep{doctorbehaviour}); language practices such as code-mixing and colloquial illness idioms (e.g., multilingual and code-mixed speech~\citep{codemixing}, language concordance, symptom-description style, local medication names~\citep{idiomsofdistress}); and emotional responses at clinical moments such as diagnosis disclosure. 
We then compared these markers against the characteristics of three datasets identified in our survey. 
\\
We analyzed 150 randomly sampled general-medicine conversations from three datasets each: ACI-Bench~\citep{yim2023aci} and MTS-Dialog~\citep{abacha2023empirical} as Western simulated references, and Eka Care Clinical Note Generation~\citep{eka_notegen}, the only publicly available simulated Indian conversational dataset. These were the closest available proxies to real clinical dialogue in our survey, since other datasets consist of single exchanges or lack conversation-level transcripts. From our broader marker set, we focus on six markers that most directly operationalise Indian--Western interactional differences (as described in Table~\ref{tab:culturalmarkers}). 
\\
ACI-Bench and MTS-Dialog both exhibit remarkably similar, clinician-dominant profiles: C:P ratios of 2.06 and 1.95, respectively (physicians produce $\simeq 2\times$ words as patients), HT:Tx ratios of 1.5 and 1.4 (treatment-oriented), and only 20\% and 18\% directive turns, with physicians instead favouring open, discursive prompts (e.g., \textit{Can you walk me through what a typical episode feels like?}). Patient agreement without follow-up occurs in just 18\% and 21\% of turns, and interruptions are effectively absent in both ($\sim$0\%). These are consistent with the structured, orderly exchanges typical of the Global North rather than the more heterogeneous interactions in the Indian clinical communication scenario.

Clinical Note Generation (Eka Care) exhibits a markedly different patient-dominant profile (C:P ratio of 0.89), with substantially more history-taking relative to treatment (HT:Tx = 4.2), more directive clinician turns (62\%), and higher patient agreement without follow-ups (64\%). Interruption rates are high (12\%, commonplace in high-volume clinical environments), unlike the near-zero rates in the other two datasets. \blu{These characteristics indicate a substantially different interactional structure between the Indian and Global North contexts, making them less representative and potentially unsuitable for evaluating Indian clinical consultations. Our hypothesis on the expected performance disparity when using translated versions of Global North datasets is supported by prior work --~\cite{koenecke2025perspective} argue that scribe errors disproportionately affect speakers with non-native accents;~\cite{hasheminasab2026evaluating} show that geography-specific disease and drug names may be absent from the vocabulary of Western models and~\cite{vaez2025interpreting} show that distribution shift in clinical text degrades model performance.} We note here that the Eka Care dataset is also simulated and not fully representative of real Indian encounters, which may reflect the characteristics of constructed conversations.

\section{How are ACS developed and deployed in India?}
We found that publicly available datasets are limited and not representative of the linguistic and interactional conditions relevant to ACS deployment in India. We interviewed individuals from organizations that develop and deploy ACS systems in India to understand how they address these limitations in practice. We conducted semi-structured interviews with representatives from four organisations (three male, one female; all graduates, with three holding postgraduate degrees and 10+ years of industry experience; demographic details in Appendix). From the interviews, we found that ACS systems are already being piloted and deployed across diverse Indian clinical settings. These deployments are largely supported by organization-specific data collection and evaluation infrastructure. 

\noindent \textbf{Diverse clinical settings:} Our participants stated wide-scale ACS deployments across settings. For instance, P1 ($\sim$2,500) and P2 (1,100) report deployments in thousands of hospitals and clinics across India. Participants also discussed optimising their ACS for the noisy background environments (P1 stated -- \textit{``$\cdots$ but half the time the quality of the sound itself was bad or with too much noise.''}), triadic consultations (P2) and regular interruptions (P3). These observations confirm that Indian clinical consultations are performed in highly challenging environments.
\\
\noindent \textbf{Localized data pipelines:}
We found that each organization independently constructs or curates datasets for training and evaluation. P3 reported that their organization collected real-world medical conversations using internal and externally sourced data for more than 50 languages through government and private grants. Similarly, P4's organization collected data by recording clinical sessions through collaborations with medical colleges and crowdsourcing through competitions. They also augmented their dataset using language models and text-to-speech systems.
P1 estimated that $\sim$75\% of their organization's 120 million patient-record corpus was India-specific, and reported support for $\sim$eight languages, including mixed-language conversations. P2 reported maintaining internal datasets collected over the past three years, with explicit attention to language-specific and code-switched clinical speech; internally fine-tuned models were more limited in language coverage, with Hindi and Malayalam-English support specifically mentioned.
\\
\noindent \textbf{Deployment-specific evaluations:} All organizations described evaluation practices that go beyond standard ASR or LLM-as-a-judge metrics, towards clinician-anchored ground truth. P3 stated: \textit{``$\cdots$ we look at a population in a particular format, then Keyword Error Rate (KER) and Keyword Recognition Rate (KRR) are more relevant.''} for evaluating whether clinically important keywords are correctly spelt and recognized, respectively. P2's organization, on the other hand, preferred speed and overall system cost: \textit{``so cost and time is the biggest
things that we generally track for $\cdots$''}. They also commented on the lack of value added by LLM-as-a-judge evaluations, since each use case is unique. Similarly, P4 also highlighted the difficulty of evaluating medical speech using generic semantic metrics: ``\textit{an LLM-based intent evaluator was found to miss important information such as dosage, drug names, and spoken representations of numerical values}''. \\
All organizations employ clinicians as part of the evaluation. P1 stated that during deployment, physician corrections were retained as feedback. For example, when a medication was transcribed incorrectly, the corrected value could be logged and tracked. Performance was also assessed on batches of consultations, with clinicians rating system outputs and documenting required edits. Similarly, P3 reported storing both generated and clinician-edited outputs to support iterative model improvement, subject to customer agreements and privacy constraints.

Thus, organizations are adapting to similar challenges independently and in parallel, but without a shared evaluation framework. ACS deployment remains highly localized across organizations, even within the same geographical, linguistic, and sociocultural context. \blu{While we cannot claim that our observations generalize, they are informative of common industry practices.} 
Next, we outline the key elements such a benchmark should include, how it should be designed, and the protocols needed to support its validity, comparability, and broad adoption.
\section{Discussion}
Our study reveals a disconnect between the deployment of ambient clinical scribes (ACS) in India and the availability of evaluation benchmarks to assess them. Our literature survey shows that publicly available datasets are fragmented across languages, modalities, provenance, and tasks, and that there is no single large-scale, real-world, multilingual, end-to-end annotated dataset for comprehensive ACS evaluation in India. 
Moreover, existing benchmarks often target a subset of related tasks rather than end-to-end evaluation, typically using simulated consultations. Such datasets offer limited insight into overall performance. \blu{While prior work has shown that scribe errors disproportionately affect speakers with non-native accents~\citep{koenecke2025perspective}, with geography-specific disease and drug names absent from the vocabulary of Western models~\citep{hasheminasab2026evaluating} and distribution shift degrading performance~\citep{vaez2025interpreting}, our quantitative analysis further establishes that Global North datasets lack the linguistic and sociocultural markers present in Indian patient-physician conversations with ACI-Bench, MTS-Dialog, and Clinical Note Generation (Eka Care) diverging across almost all metrics.} An evaluation suite must therefore also be annotated with linguistic and sociocultural markers. Finally, our interviews show that this gap does not prevent deployment with organizations present across heterogeneous healthcare settings in India. They are independently collecting proprietary local data and adapting their systems and evaluation suites to Indian deployment settings \blu{-- keyword-level metrics for P3; cost and time for P2; clinician ratings and edits for P1; thus, their claims cannot be compared.} This has made capabilities difficult to compare against a common reference, increasing the risk of biases and context-specific failure modes going undetected until deployment. The absence of a shared, publicly available evaluation benchmark has shifted the burden of establishing the validity of individual vendor claims onto healthcare systems and regulators. 

\subsection{A benchmark for Indian ACS}
Our findings point to four properties a useful Indian ACS benchmark should embody:\\
\noindent $\bullet$ \textbf{Preference for real data:} Real consultations should be primary and can be augmented with synthetic or simulated data, especially for low-resource languages or rare specialties.\\
\noindent $\bullet$ \textbf{Regional linguistic diversity:} It should reflect the deployment region's languages, accents, and code-mixed speech directly, not via translation.\\
\noindent $\bullet$ \textbf{Representative variation:} It should capture diversity of language, speaker demographics and role, count, non-PII sociodemographic attributes such as gender and age (with consent), background noise (in dB) and healthcare settings.\\
\noindent $\bullet$ \textbf{Clinically relevant metrics:} Clinically relevant evaluation metrics and dimensions should include not only conventional measures such as WER for ASR and F1 for clinical NER, but also clinically meaningful measures of information preservation, speaker attribution, terminology accuracy, omissions, hallucinations, structured-field accuracy, and human-in-the-loop review, along with fairness-appropriate metrics like equalized odds~\citep{hardt2016equality} and equal opportunity~\citep{mehrabi2021survey}.

The benchmark should offer common reference tasks for cross-system comparison-- standard evaluations combined with targeted tasks such as speaker attribution, noise detection, and medical terminology identification, enabling comparability without disclosing proprietary data or models. This would give regulators and health systems a common baseline for assessing safety, reliability, and failure modes. Even though such benchmarks are hard to construct due to a lack of real consultations (major concerns around consent, ownership, anonymization, and sharing), approaches like a consortium-based model -- where data are contributed under shared governance and only controlled evaluation subsets are released, supplemented by synthetic or simulated data for low-resource languages and rare specialties -- may offer a path forward.

\subsection{Limitations and future work}
Our study has three main limitations. The survey covers only publicly released datasets, omitting undocumented proprietary ones. Our sociocultural representativeness assessment is bounded by what the dataset authors report. And our interview analysis, drawn from four organizations' self-reported accounts, may not capture the full diversity of ACS developers and deployments in India; we treat these accounts as evidence of current practice rather than independent measures of system quality. Future work will extend our cultural-marker analysis, develop a representative ACS benchmark, and evaluate deployed systems against it.

\bibliography{main}
\clearpage
\begin{table*}[!ht]
    \centering
    \scriptsize
    \setlength{\tabcolsep}{2pt}
        \caption{Demographic and professional characteristics of the four industry interview participants. Participants are anonymized as P1--P4; N.A. indicates information not available or not reported.}
    \label{tab:interviewparticipants}
    \begin{tabular}{cccccc}
    \toprule
        \textbf{Alias} & \textbf{Gender} & \textbf{Age} & \textbf{Highest Education} & \textbf{Relevant experience} & \textbf{Role}\\
        \midrule
            P1 & Male & N.A. &  MBA & 17 years in health-tech & Co-founder \\ 
            P2 & Male & 30 & Engg. graduate & 10+ years software engineering; ~8 years open source & CEO \\ 
            P3 & Female & N.A. & Ph.D. & $\sim$ 4.5 years working in V2DD/medical AI & Founder \& CEO \\ 
            P4 & Male & 40 &  Ph.D. & $\sim$18–19 years in V2DD medical AI  & Co-founding team member \\ 
    \bottomrule
    \end{tabular}
\end{table*}

\appendix
\section{Quantitative Method Details}\label{appendix:quant}
\subsection{Dataset Details}
\blu{We used 150 samples from each dataset because Eka Care Clinical Note Generation~\cite{eka_notegen}, the smallest of the three datasets used in the analysis, contains 156 samples. We randomly sampled 150 first-time visits from each dataset. To improve comparability, we restricted the reference datasets to general outpatient consultations, rather than specialist or emergency encounters, and stratified them by patient gender, where available. Transcript lengths are naturally variable, including within the same dataset, as is typical of general consultations. While these criteria do not eliminate all differences across datasets, we use the filtering for first-time consultations and general outpatient encounters to provide a comparable basis for our conversational analysis.}

\subsection{Cultural marker framework}
\blu{Our quantitative evaluation employs a multilevel interactional cultural marker framework. At the first level -- $L0$ (transcript level), markers such as word count, turn-length uniformity, backchannel rate, and vocative frequency are computed directly from the transcript. The second level -- $L1$ (inter-turn level) captures interactional dynamics such as interruption, topic continuity, echo, and related inter-turn behaviours, while the deepest level -- $L2$ (turn level) captures properties such as doctor directiveness, register, verbosity, question structure, patient response style, symptom framing, information volunteering, and agreement/reassurance style. A conditional physical-examination layer is applied when an examination is present. These dimensions were derived from the interactional patterns identified in prior work and refined through feedback from two medical doctors to ensure clinical relevance.} 

\subsection{Metrics for evaluation}
\blu{Each quantity presented in Table~\ref{tab:culturalmarkers} is computed as follows--C:P ratio is computed from the total number of words produced by the clinician and patient and computes the clinician-to-patient word ratio. A value below 1 indicates that patients contribute more words overall. HT:Tx ratio is the number of history-taking turns divided by the number of treatment turns. Turns are first assigned to the relevant consultation phases inspired by Byrne and Long's consultation model~\citep{Gray188}, allowing us to distinguish history-taking from treatment. Directiveness is derived from the L3 Tone annotation. Each clinician turn is classified as directive, discussive, or empathetic. The reported percentage is the proportion of clinician turns classified as directive. Patient turns containing symptom descriptions are classified according to their linguistic framing (clinical, colloquial, or deictic). The Table~\ref{tab:culturalmarkers} value reports the proportion of patient turns containing medical/technical vocabulary, i.e., the clinical form of symptom description. Response-Agreement is derived from classifying patient responses as accepting, questioning, expressing concern, or refusing. The reported value is the proportion of patient turns that accept or agree with the clinician's preceding contribution without asking a follow-up question.
Interruptions measures how often the clinician cuts off or redirects an ongoing patient response before the patient’s narrative is complete. All these measures are computed for each turn and averaged across the entire dataset.}

\subsection{Details of LLM-as-a-judge}
\blu{For the LLM-based annotation, we use GPT-5.1 with a temperature of $0$. The model is given the relevant turn or turn pair and is prompted to assign the appropriate category from the proposed annotation scheme. To validate the LLM-based annotation, we conduct human annotation using the same layer-by-layer scheme. We sample 10 conversations per language and have two annotators independently apply the framework. The resulting human-human agreement is substantial, with Cohen’s $\kappa$ of 0.71, 0.68, and 0.72 for English, Hindi, and Telugu, respectively. Agreement between the averaged human annotations and the LLM annotations is 0.846, 0.813, and 0.804, respectively.}

\section{Interview Participant Details}\label{apd:part_details}
In Table~\ref{tab:interviewparticipants}, we provide the detailed demographic details of our interview participants from four different organizations developing ambient clinical scribes (ACS) in India and deploying in India.




\end{document}